\documentclass[aps,twocolumn,secnumarabic,nobalancelastpage,nofootinbib, showpacs]{revtex4}
\usepackage{amsmath,amsfonts,amssymb}
\usepackage{wrapfig}
\usepackage{subfigure}
\usepackage{graphicx}
\usepackage{bbm}

\usepackage{color}
\usepackage{verbatim}

\def \beq {\begin{equation}}
\def \eeq {\end{equation}}
\def \ba {\begin{eqnarray}}
\def \ea {\end{eqnarray}}
\def \l {\left}
\def \r {\right}

\newcommand{\Mean}[1]{\left\langle#1\right\rangle}
\newcommand{\pd}[2]{\frac{\partial #1}{\partial #2}}

\newcommand{\nn}{\nonumber}
\newcommand{\mc}{\mathcal}

\def\ket#1{\left| #1\right>}

\newcommand{\bas}{\begin{eqnarray*}}
\newcommand{\eas}{\end{eqnarray*}}

\begin{document}

\title{Dynamics of an Ion Coupled to a Parametric Superconducting Circuit}
\author{Dvir Kafri$^1$, Prabin Adhikari$^1$,  Jacob M. Taylor$^{1,2}$}
\affiliation{$^1$Joint Quantum Institute, University of Maryland, College Park}
\affiliation{$^2$National Institute of Standards and Technology, Gaithersburg, Maryland}

\begin{abstract}

Superconducting circuits and trapped ions are promising architectures for quantum information processing. However, the natural frequencies for controlling these systems -- radio frequency ion control and microwave domain superconducting qubit control -- make direct Hamiltonian interactions between them weak. In this paper we describe a technique for coupling a trapped ion's motion to the fundamental mode of a superconducting circuit, by applying to the circuit a carefully modulated external magnetic flux. In conjunction with a non-linear element (Josephson junction), this gives the circuit an effective time-dependent inductance. We then show how to tune the external flux to generate a resonant coupling between the circuit and ion's motional mode, and discuss the limitations of this approach compared to using a time-dependent capacitance.

\end{abstract}


\maketitle

\section{Introduction}

Superconducting circuits and trapped ions have distinct advantages in quantum information processing. Circuits are known for fast gate times, flexible fabrication methods, and macroscopic sizes, allowing multiple applications in quantum information science\cite{Dicarlo2009, Clarke2004, Blais2004, Zakka-Bajjani2011}. Unfortunately, they have short coherence times and their decoherence mechanisms are hard to address\cite{Devoret2013}. Trapped ions, on the other hand, serve as ideal quantum memories. Indeed, the hyperfine transition displays coherence times on the order of seconds to minutes\cite{Bollinger1991,Roos2004,Langer2005,Harty2014}, while high fidelity state readout is available through fluorescence spectroscopy \cite{Leibfried2003,Dehmelt1975}. Unfortunately, ions depend mainly on motional gates for interactions \cite{Cirac1995, Molmer1999, Blatt2008, Monz2009}; these are correspondingly slow, susceptible to motional heating associated with traps\cite{Wineland1998,Turchette2000,Deslauriers2004}, and occur only at a short -- dipolar -- range. The distinct advantages of ion and superconducting systems therefore motivates a hybrid system comprising both architectures, producing a long-range coupling between high-quality quantum memories.

Early proposals for hybrid atomic and solid state systems \cite{Sorensen2004,Tian2004} have yet to be implemented experimentally. Other approaches involve coupling solid state systems to atomic systems with large dipole moments, such as ensembles of polar molecules\cite{Rabl2006} or Rydberg atoms\cite{Petrosyan2009}. Although the motional dipole couplings with the electric field of superconducting circuits can be several hundred $\rm{kHz}$, these systems suffer from a large mismatch between motional ($\sim \rm{MHz}$) and circuit ($\sim \rm{GHz}$) frequencies. This causes the normal modes of the coupled systems to be either predominantly motional or photonic in nature, thereby limiting the rate at which information is carried between them. Implementation of a practical hybrid device therefore requires something additional.

Parametric processes allow for efficient conversion of excitations between off-resonant systems. In the field of quantum optics, they are widely used in the frequency conversion of photons using nonlinear media \cite{Burnham1970, Mandel1995}. In the realm of superconducting quantum devices, parametric amplifiers provide highly sensitive, continuous readout measurements while adding little noise \cite{Bergeal2010,Castellanos-Beltran2008,Vijay2009,Vijay2011}.  Parametric processes can also been used to generate controllable interactions between superconducting qubits and microwave resonators \cite{Beaudoin2012,Strand2013,Allman2014}. In the context of hybrid systems, Ref.~\cite{Kielpinski2012} presents a parametric coupling scheme between the resonant modes of an LC circuit and trapped ion. The ion, confined in a trap with frequency $\omega_{i}$, is coupled to the driven sidebands of a high quality factor parametric LC circuit whose capacitance is modulated at frequency $\omega_{LC} - \omega_{i}$. This gives rise to a coupling strength on the order of tens of kHz for typical ion chip-trap parameters. A different approach proposed in Ref.~\cite{Daniilidis2013} is based on a position-charge interaction that is quadratic in the charged particle's position, so that driving of its motion produces a parametric coupling. In this approach the circuit couples to an electron rather than an ion, leading to an enhanced coupling strength ($\sim 1$ MHz) due to its reduced mass.

In this manuscript we describe an alternative parametric driving scheme to produce coherent interactions between atoms and circuits. By using a time-dependent external flux, we drive a superconducting loop containing a Josephson junction, which causes the superconductor to act as a parametric oscillator with a tunable inductance. By studying the characteristic, time-dependent excitations of this system, we show how to produce a resonant interaction between it and the motional mode of a capacitively coupled trapped ion. Although in principle our approach could be used to produce a strong coupling, we find that, in contrast with capacitive driving schemes, the mismatch between inductive driving and capacitive (charge-mediated) interaction causes a significant loss in coupling strength.  

The manuscript proceeds as follows: In the first section, we describe the experimental setup of the circuit and ion systems and motivate the method used to couple them. To understand how the Josephson non-linearity and external flux affect the circuit, we transform to a reference frame corresponding to the classical solution of its non-linear Hamiltonian. We then linearize about this solution, resulting in a Hamiltonian that describes fluctuations about the classical equations of motion and corresponds to an LC circuit with a sinusoidally varying inductance. This periodic, linear Hamiltonian is characterized by the quasi-periodic solutions to Mathieu's equation. From these functions we define the `quasi-energy' annihilation operator for the system and transform to a second reference frame where it is time-independent. Finally, we derive the circuit-ion interaction in the interaction picture, which allows us to directly compute the effective coherent coupling strength between the systems. We conclude by comparing our results to previous work\cite{Kielpinski2012} (a capacitive driving scheme) and analyzing why inductive modulation is generically ineffective for capacitive couplings.

\section{Physical System and Hamiltonian}

We begin with a basic physical description of the ion-circuit system, and as in Ref.~\cite{Devoret1995}, construct the classical Lagrangian before deriving the quantized Hamiltonian. We consider a single ion placed close to two capacitive plates of a superconducting circuit. We assume that the ion is trapped in an effective harmonic potential with a characteristic frequency $\omega_z$ in the direction parallel to the electric field between the two plates\cite{Leibfried2003}. The associated ion Lagrangian is 
\beq
\mathcal{L}_{ion}(z, \dot z, t) = \frac{1}{2} m \dot z^2  - \frac{1}{2} m \omega_z^2 z^2 \,,
\eeq
where $m$ is the ion mass and $z$ its displacement from equilibrium. Since the LC circuit will only couple to the ion motion in the $z$ direction, we ignore the Lagrangian terms associated with motion in the other axial directions.

\begin{figure}[t]
\centering 
\includegraphics[width=6cm]{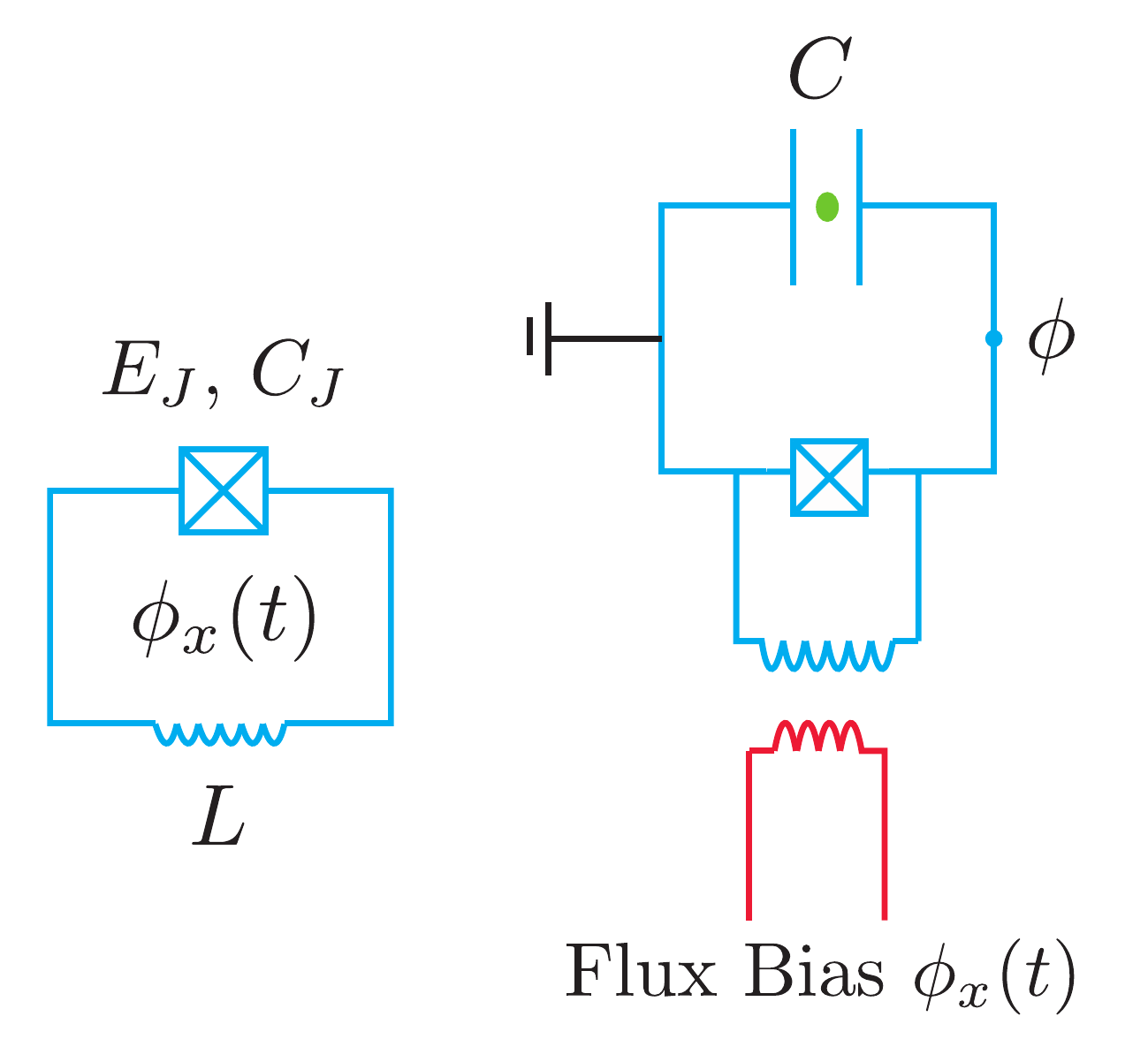} 
\caption{Left: An rf-Squid with Josephson energy $E_{J}$ and junction capacitance $C_{J}$ driven by a time-dependent external flux $\phi_{x}(t)$. The outer loop with inductance $L$ contributes an energy $E_{L}$. Right: Schematic of an ion confined by a capacitor with capacitance $C$. The capacitor is connected in parallel to the rf-SQUID system.}
\label{Circuit}
\end{figure}

The circuit interacting with the ion contains a Josephson junction \cite{Likharev1986} (Fig.~\ref{Circuit}) shunted by an inductive outer loop with inductance $L$. This configuration is known as a radio frequency superconducting quantum interference device (rf-SQUID)\cite{Clarke2004}. The SQUID is connected in parallel to a capacitor $C$, whose plates are coupled to the nearby trapped ion. An external, time-dependent magnetic flux $\phi_{x}(t)$ through the circuit is used to tune the characteristic frequencies of this system. In our proposed configuration, the SQUID Lagrangian can be written in terms of the node flux $\phi$ as\cite{Devoret1995}
\beq
\label{Lsquid}
\mathcal{L}_{q}(\phi,\dot \phi, t) = \frac{1}{2} C_\Sigma \dot{\phi}^2 +  E_{J} \cos ( \phi/\tilde \phi_0) - \frac{1}{2 L}  (\phi-\phi_{x}(t))^2  \,,
\eeq
where $\tilde \phi_0 = \hbar/(2 e)$ is the reduced flux quantum, and $C_\Sigma = C + C_J$ is the effective capacitance of the circuit including the capacitance $C_J$ of the Josephson junction. The Josephson energy is defined in terms of the critical current of the junction, $E_J = I_c \tilde \phi_0$. The Josephson term adds a non-linearity to the system $\propto E_J$ which will, in conjunction with the external driving, allow for coherent transfer of excitations between the otherwise off-resonant LC circuit (typical frequency $\sim 1-10 \mbox{ GHz}$) and ion motional mode ($\sim 1-10 \mbox{ MHz}$). Finally, we include the interaction between the two systems, associated with the ion motion through the electric field between the two plates in the dipole approximation:
\ba
\mathcal{L}_I =  -e \l(\frac{\xi}{d} \dot{\phi}\r) z \,,
\ea
where $d$ is the distance between the two plates and $\xi\sim 0.25$ a dimensionless factor associated with the capacitor geometry\cite{Kielpinski2012}. 

With the full Lagrangian $\mc{L} = \mc{L}_q + \mc{L}_{ion} + \mc{L}_I$ we follow the standard prescription to define the canonical variables conjugate to $z$ and $\phi$:
\ba
\label{p}
p_{z} &=& \pd{\mc{L}}{\dot{z}} = m \dot{z}\\
\label{q} q &=& \pd{\mc{L}}{\dot{\phi}} = C_{\Sigma} \dot{\phi} - \xi \frac{e}{d} z\,.
\ea
These are identified as the momentum of the ion in the $z$ direction and the effective Cooper-pair charge difference across the Josephson junction, respectively. We use the canonical Legendre transformation $H = q \dot \phi + p_z \dot z - \mc{L}$ to define the Hamiltonian and quantize the system, giving
\beq
\hat H(t) = \hat H_{ion} + \hat H_{q}(t) + \hat H_{I}\,,
\eeq
where 
\ba
\hat H_{ion} &=& \frac{\hat p_{z}^2}{2m} + \frac{1}{2} m  \omega_{i}^2 \hat z^2, \\
\hat H_{q}(t) &=& \frac{\hat q^2}{2 C_{\Sigma}} -E_{J} \cos(\hat \phi/\tilde \phi_0) + \frac{1}{2 L}  (\hat \phi-\phi_{x}(t))^2, \\
\hat H_{I} &=&  e\frac{ \xi}{d C_{\Sigma}} \hat q \hat z\,,
\ea
As the canonical charge $q$ of equation \eqref{q} has a term proportional to $z$, the ion Hamiltonian $\hat H_{ion}$ gains an extra potential term proportional to $\hat z^2$. Accounting for this term, we define the dressed trap frequency, 
\ba
\omega_{i}^2 = \omega_{z}^2 + \frac{e^2 \xi^2}{d^2 C_\Sigma m}\,.
\ea
The latter correction is small compared to $\omega_z$; it is on the order of $50 \mbox{kHz}$ for the parameters suggested in Ref.~\cite{Kielpinski2012} ($d \approx 25 \mu\mbox{m}$, $C_\Sigma \approx 50 \mbox{fF}$ and $m \approx 1.5 \times 10^{-26}\mbox{kg}$ the atomic mass of Beryllium). Finally, we note the resulting commutation relations, $[\hat \phi,\hat q] = [\hat z, \hat p_z] = i \hbar$. 

To motivate the method for coherently coupling the ion and circuit, we consider the Hamiltonian in the interaction picture, i.e., in the frame rotating with respect to $\hat H_q(t) + \hat H_{ion}$. In this frame the Hamiltonian takes the form, 
\beq
\label{Hint}
\hat H_{int}(t) = \frac{\xi e}{d C_\Sigma} \hat q(t) \l( \hat b e^{-i \omega_i t} + \hat b^\dagger e^{i \omega_i t}\r) \sqrt{\frac{\hbar}{2 m \omega_i}}\,,
\eeq
where $\hat b = \sqrt{\frac{m \omega_i}{2 \hbar}} \hat z + i \sqrt{\frac{1}{2 \hbar m \omega_i}} \hat p_z$ is the annihilation operator associated with excitations of the ion motional mode, and $\hat q(t)$ is propagated according to the time-dependent Hamiltonian $\hat H_{q}(t)$. Observe that if we set both $\phi_x(t)$ and $E_J$ to $0$ in the definition of $\hat H_q(t)$, the charge $\hat q(t)$ will oscillate as a harmonic oscillator, 
\beq
\hat q(t) \rightarrow i\sqrt{\frac{\hbar C_\Sigma \omega_0}{2 }}(\hat a e^{-i \omega_0t}- \hat a^\dagger e^{i \omega_0t})\,,
\eeq
with $\hat a$ the circuit annihilation operator analogous to $\hat b$, and $\omega_0$ the undriven frequency LC,
\beq
\omega_0 = 1/\sqrt{L C_\Sigma}\,.
\eeq
In order to coherently exchange excitations between the systems, $\hat H_{int}(t)$ should only contain beam splitter-like terms, $\hat a \hat b^\dagger$ and $\hat a^\dagger \hat b$, while suppressing the excitation non-conserving terms, $\hat a^\dagger \hat b^\dagger$ and $\hat a \hat b$. Yet since $\omega_0 \gg \omega_i$, the $a b^\dagger$ terms oscillate at a frequency comparable to the $a^\dagger b^\dagger$ terms, both of which have negligible effect in the rotating wave approximation (RWA). Our goal is therefore the following: design $E_J$ and $\phi_x(t)$ such that $\hat a \hat b^\dagger$ contains a time-independent component in the interaction picture, while $\hat a \hat b$ contains only oscillating terms that drop out in the RWA.

\section{Linearization of the Parametric Oscillator}
Before we can determine the parameters $\phi_x(t)$ and $E_J$ allowing for coherent exchange of excitations, we must first bring $\hat H_q(t)$ into a more agreeable form. To begin, we linearize the circuit Hamiltonian about the classical solution to its equations of motion. This makes it possible (in the next section) to identify effective annihilation and creation operators for the LC system, and to analyze their spectrum.



We linearize by displacing the charge and flux variables by time-dependent scalars, $q_c(t)$ and $\phi_c(t)$, through the unitary transformations
\ba
\nn \hat U_{1}(t) &=&  e^{i \hat \phi q_{c}(t)/\hbar}, \\
\hat U_{2}(t) &=&  e^{-i \hat q \phi_{c}(t)/\hbar}\,.
\ea
Specifically, we consider the LC circuit in terms of the displaced states, 
\beq
\ket{\tilde \Psi(t)} = \hat U^\dagger_2(t) \hat U^\dagger_1(t) \ket{\Psi(t)}\,,
\eeq
whose equation of motion satisfies
\ba
\begin{aligned}
\partial_t \ket{\tilde \Psi(t)} &= -i \hbar^{-1}\tilde H_q(t) \ket{\tilde \Psi(t)}\,,\\
\label{Hrot}
\tilde H_q &= \hat U_2^\dagger \hat U_1^\dagger \hat H_q \hat U_1 \hat U_2 - i \hbar \hat U_2^\dagger \hat U_1^\dagger \frac{\partial \hat U_1}{\partial t} \hat U_2\\
& - i \hbar \hat U_2^\dagger  \frac{\partial \hat U_2}{\partial t}\,.
\end{aligned}
\ea
Given $[\hat \phi, \hat q] = i \hbar$, we have that $\hat U_1^\dagger \hat q \hat U_1 = \hat q + q_c(t) $ and $\hat U_2^\dagger \hat \phi \hat U_2 = \hat \phi + \phi_c(t) $, while clearly $\hat U_1^\dagger \hat \phi \hat U_1 = \hat \phi$ and $\hat U_2^\dagger \hat q \hat U_2 = \hat q$. The displaced state Hamiltonian is therefore
\ba
\tilde H_q(t) &=& \frac{(\hat q+q_{c}(t))^2}{2C_{\Sigma}} + V_{q}(\hat \phi+\phi_{c}(t)) \nn \\
& &  + (\hat \phi+\phi_{c}(t)) \partial_t q_{c}(t) -\hat q \partial_t \phi_{c}(t),
\ea
where we have defined the nonlinear potential, 
\beq
\label{Vq}
V_q(\hat \phi )=  - E_J \cos( \hat \phi/\tilde \phi_0) + \frac{1}{2 L}( \hat \phi - \phi_x(t))^2 \,,
\eeq
with $\tilde \phi_0 = \hbar/(2 e)$. We now Taylor expand $V_q$ about $\phi_c(t)$, and collect all first and second order terms in $\hat q $ and $\hat \phi$, giving
\ba
\tilde H_q(t) &=& \hat q \l( q_c(t)/C_\Sigma - \partial_t \phi_c \r) + \hat \phi \l( V_q'(\phi_c) + \partial_t q_c \r)\nn\\
&& + \frac{\hat q^2}{2 C_\Sigma} + \frac{V_q''(\phi_c)}{2!}\hat \phi^2 + R(\hat \phi)\,,
\ea
where we have dropped all scalar terms, and 
\beq
\label{R}
R(\hat \phi) = \sum_{k \geq 3} \tilde \phi_0^k V_q^{(k)}(\phi_c)\l(\hat \phi/\tilde \phi_0\r)^k\frac{1}{k!}
\eeq
represents all higher order terms in the Taylor expansion. 

To complete the linearization, the displacements $q_c$ and $\phi_c$ are chosen so that the first order terms in $\hat q$ and $\hat \phi$ vanish, and therefore $\tilde H_q$ is quadratic to leading order:
\ba
\partial_t \phi_c &=& q_c/C_\Sigma\nn\\
\label{classical}\partial_t q_c &=& - V_q'(\phi_c)\,.
\ea
These relations are the solutions to the classical, driven Hamiltonian, $H_q = \frac{q_c^2}{2 C_\Sigma} + V_q(\phi_c,t)$, and can be substituted into each other to give
\beq
\label{driving}
\partial_t^2 \phi_c  + \omega_0^2 (\phi_c + \beta \tilde \phi_0 \sin(\phi_c/\tilde \phi_0)) = \omega_0^2 \phi_x(t) \,,
\eeq
where
\beq
\beta = L E_J/\tilde \phi_0^2 = \frac{L I_c}{\tilde \phi_0}
\eeq 
represents the strength of the non-linearity and $\omega_0 = 1/\sqrt{L C_\Sigma}$ is the bare LC resonance frequency. Substituting from \eqref{classical} and computing $V_q''(\phi_c) = \frac{1}{L}(1 + \beta \cos(\phi_c/\tilde \phi_0))$, we can now express the circuit Hamiltonian as
\beq
\label{tHq}
\tilde H_q(t) = \frac{ \hat q^2}{2 C_\Sigma} + \frac{1}{2 L } (1 + \beta \cos(\phi_c(t)/\tilde \phi_0)) \hat \phi^2 \,.
\eeq
Note that we have dropped the higher order terms $R(\hat \phi)$ of equation~\eqref{R}. Understanding when this is valid will require us to express the flux operator $\hat \phi$ in the interaction picture, which we carry out below. A complete analysis is deferred to the appendix.

\begin{figure}
\includegraphics[width=.45\textwidth]{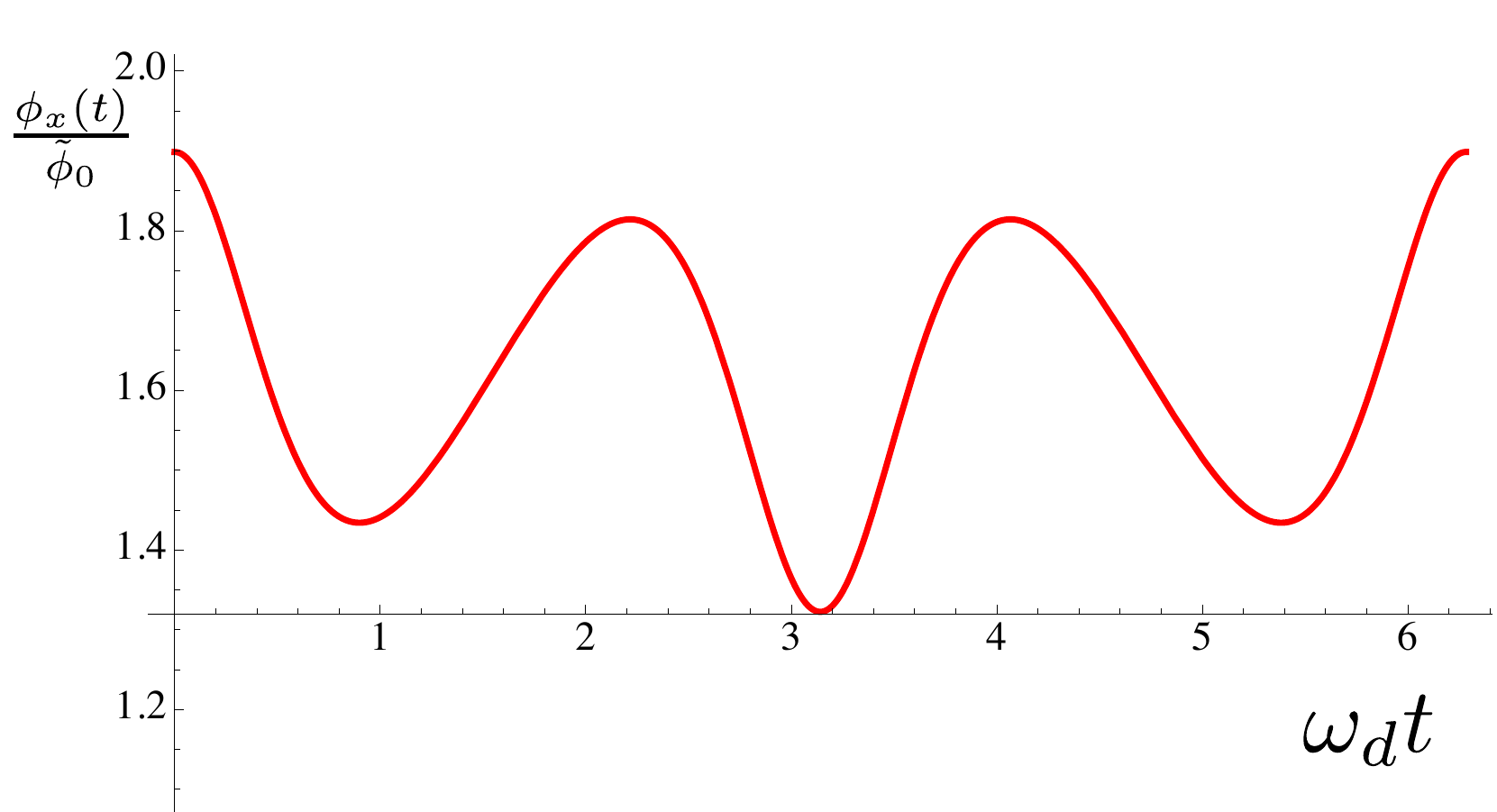} 
\caption{ External flux drive $\phi_x(t)/\tilde \phi_0$ required for a sinusoidal modulation of circuit inductance. The flux $\phi_x(t)$ is defined explicitly by equation \eqref{driving}, where the classical solution $\phi_c(t)$ satisfies $\beta \cos(\phi_c(t)/\tilde \phi_0) = \eta  \cos(\omega_d t)$. The plotted function corresponds to drive parameters $\omega_d = 0.999 \omega_0$, $\eta = 0.06 = (3/4)\times \beta$, where $\beta = E_J (\tilde \phi_0^2/L)^{-1}$ represents the  strength of the non-linearity compared to the bare linear inductance. The relative scaling $\eta/\beta<1$ is chosen so that $\phi_c$ is a smooth function of time; the $\eta\rightarrow \beta^-$ limit corresponds to a saw-tooth wave with undefined second derivative. }
\label{fluxDrive}
\end{figure}

With equation \eqref{tHq} we have converted to the Hamiltonian of a harmonic oscillator with time-dependent inductance. Although it is explicitly dependent on the classical solution $\phi_c(t)$, we note that engineering this function is rather straightforward. Given a desired $\phi_c(t) $, the external driving $\phi_x(t)$ needed to produce it is given explicitly by equation \eqref{driving} (see Fig.~\ref{fluxDrive}). For simplicity we assume that $\phi_c(t)$ satisfies the relation
\beq
\beta \cos(\phi_c(t)/\tilde \phi_0) = \eta \cos(\omega_d t)\,,
\eeq
where we assume $\eta<\beta$ so that $\phi_c$ is well defined at all $t$. This corresponds to an LC circuit with inverse inductance $L^{-1}$ modulated at amplitude $\eta$ and frequency $\omega_d$,
\beq
\label{tHq2}
\tilde H_q(t) = \frac{ \hat q^2}{2 C_\Sigma} + \frac{1}{2 L } (1 + \eta \cos(\omega_d t)) \hat \phi^2 \,.
\eeq 
In the following section we will see how the design parameters $\eta$ and $\omega_d$ determine the evolution of this system.

\section{Time-Dependent Quantum Harmonic Oscillator}
\label{TDQHO}

Although we now have a quadratic Hamiltonian for the SQUID, since $\tilde H_q(t) = {\frac{\hat q^2}{2 C_\Sigma} + \frac{1}{2 L }(1 + \eta \cos(\omega_d t)) \hat \phi^2}$ is time-dependent, it is not immediately clear how to define its associated annihilation operator. To do so, we follow the approach of Ref.~\cite{Brown1991} to obtain an effective operator $\hat a$ retaining useful properties of the time-independent case. Specifically, it satisfies $[\hat a, \hat a^\dagger] = 1$ and, in the appropriate reference frame, is explicitly time-independent. Further, in this frame the Hamiltonian can be written as $\hbar (\partial_t \theta(t))( \hat a^\dagger \hat a + 1/2)$, where $\theta(t)$ is the effective phase accumulated by $\hat a$ in the Heisenberg picture. This new Hamiltonian commutes with itself at all times, allowing us to directly transform to the interaction picture in the next section. We will then describe how to control the spectral properties of $\hat q(t)$ in the interaction Hamiltonian \eqref{Hint}, producing the desired time-independent beam splitter-like interaction between circuit and ion motion.

Following Ref.~\cite{Brown1991}, we use unitary transformations to change $\tilde H_q(t)$ so that it is of the form $\sim g(t) ( \alpha \hat q^2 + \delta \hat \phi^2 )$, where $\alpha$ and $\delta$ are explicitly time-independent. To do this, we first analyze the classical equation of motion for the flux $\phi$ associated with $\tilde H_q(t)$, as derived from equation \eqref{tHq2},
\beq
\label{MathieuFirst}
\partial_t^2 f(t) + \omega_0^2(1 + \eta \cos(\omega_d t)) f(t) = 0\,.
\eeq
Since the drive has period $\tau = 2 \pi/\omega_d$, from Floquet's theorem \cite{Magnus1979,Kelley2010} we can find a quasi-periodic solution $f$ satisfying
\ba
\nn f(0) &=& 1\,,\\
\label{fPeriod}
 f(t + \tau) &=& e^{i \mu \pi} f(t)\,.
\ea
In order for the solution to be stable, the characteristic exponent $\mu$ must be real valued. This imposes constraints on the parameters $\omega_d$ and $\eta$, which we discuss later. As we shall see, the properties of the function $f$ will be closely related to the spectrum of $\hat q(t)$ in the interaction picture.  

It is useful to express $f$ in polar form,
\beq
f(t) = r(t) e^{i \theta(t)}\,,
\eeq
where $r>0$ and $\theta$ is real valued. Substituting this relation into the characteristic equation \eqref{MathieuFirst} determines equations of motion for $r$ and $\theta$,
\ba
\label{polarHill}
\begin{split}
\partial_t^2  r & = \l( (\partial_t \theta)^2-\omega_0^2(1 + \eta \cos(\omega_d t))\r) r \,, \\
0 &= r \partial_t^2 \theta_t + 2 (\partial_t r) (\partial_t \theta)\,.
\end{split}
\ea
We define the associated Wronskian, which sets a characteristic frequency scale for the evolution,
\ba
\nn W &=& \frac{1}{2 i}\l(f^*(t) \partial_t f(t) - f(t) \partial_t f^*(t)\r) \\
\nn & =& \mbox{Im}\l( r e^{-i \theta} \partial_t \l(r e^{i \theta} \r)  \r)\\ 
\label{Wronskian} & = & r^2 \partial_t \theta \,.
\ea 
We note that $\frac{1}{r}\partial_t W$ is equal to the second line of \eqref{polarHill}, so $\partial_t W = 0$ and $W$ is time-independent. 

The above definitions are key to the transformations bringing $\tilde H_{q}(t)$ into the desired form. Our first transformation is
\beq
\hat U_3 = \exp(i \chi(t) \hat \phi^2/\hbar)\,,
\eeq
where
\beq
\chi(t) \equiv \frac{C_\Sigma}{2}\frac{\partial_t r}{r}
\eeq
is the real part of the effective admittance, $\frac{C_\Sigma}{2} \frac{\partial_t f}{ f}$. Using $[ \hat \phi, \hat q] = i \hbar$, we note
\ba
\nn \hat U_3^\dagger \hat q U_3 &=& \hat q + 2 \chi \hat \phi\\
 \nn\hat U_3^\dagger \hat \phi U_3 &=& \hat \phi \\
-i \hbar \hat U_3^\dagger \partial_t \hat U_3 &=&  (\partial_t \chi) \hat \phi^2
\ea
After some algebra, the Hamiltonian of Eq. \eqref{tHq2} is transformed to $\hat U_3^\dag \tilde H_q(t) \hat U_3 - i \hbar \hat U_3^\dagger \partial_t \hat U_3$ and becomes
\ba
\nn  && \frac{\hat q^2}{2 C_\Sigma} + \frac{ \chi}{C_\Sigma} (\hat \phi \hat q + \hat q \hat \phi)\\
&& +\frac{C_\Sigma \hat \phi^2}{2}\l(  \omega_0^2\l( 1 + \eta \cos(\omega_d t) \r) +  (2 \chi/C_\Sigma)^2 + 2 \partial_t \chi/C_\Sigma \r) \\
\nn & = & \frac{\hat q^2}{2 C_\Sigma} + \frac{ \chi}{C_\Sigma} (\hat \phi \hat q + \hat q \hat \phi) + \frac{C_\Sigma}{2} (\partial_t \theta)^2 \hat \phi^2\,.
\ea
In the last line we used $2 \chi/C_\Sigma =(\partial_t r)/r$ and the first line of \eqref{polarHill} to compute $2 \partial_t \chi/C_\Sigma = (\partial_t \theta)^2 - \omega_0^2\l( 1 + \eta \cos(\omega_d t) \r) - (2 \chi/C_\Sigma)^2$.  

The next transformation removes the cross term and rescales $\hat q$ and $\hat \phi$:
\beq
\hat U_4 = e^{-i F(t) (\hat \phi \hat q + \hat q \hat \phi)/\hbar} \,,
\eeq
where $F(t) = \frac{1}{2} \log(r)$ satisfies $\partial_t F =  \chi/C_\Sigma$. Given $[(\hat \phi \hat q + \hat q \hat \phi) , \hat \phi] =- 2 i \hbar \hat \phi$ and $[(\hat \phi \hat q + \hat q \hat \phi) , \hat q] = 2 i \hbar \hat q$, we compute
\ba
\nn \hat U_4^\dagger \hat q U_4 &=&  e^{-2 F} \hat q = \frac{\hat q}{r} \\
\nn \hat U_4^\dagger \hat \phi U_4 &=& e^{2 F} \hat \phi =  r \hat \phi  \\
-i \hbar \hat U_4^\dagger \partial_t \hat U_4 &=& -\frac{ \chi}{C_\Sigma} (\hat \phi \hat q + \hat q \hat \phi)\,.
\ea
The final transformed Hamiltonian is therefore
\ba
\nn \hat H_{LC} & = & \frac{\hat q^2}{2 C_\Sigma r^2} + \frac{C_\Sigma}{2} ( \partial_t \theta)^2 r^2 \hat \phi^2\\
\label{HLC} & = & \frac{\partial_t \theta }{W} \l(\frac{\hat q^2}{2 C_\Sigma } + \frac{1}{2} C_\Sigma W^2 \hat \phi^2 \r) \,,
\ea
where the last line follows from equation \eqref{Wronskian}.

Equation \eqref{HLC} allows us to define the effective annihilation operator in the standard way,
\beq
\label{a}
\hat a = \sqrt{\frac{ C_\Sigma W}{2 \hbar}} \hat \phi + i \sqrt{ \frac{1}{2\hbar C_\Sigma W}} \hat q\,,
\eeq
so that the Hamiltonian is also equal to
\beq
\label{HLCtrans}
\hat H_{LC}(t) =  \hbar (\partial_t \theta) (\hat a^\dagger \hat a + 1/2) \,.
\eeq
In this reference frame, we can interpret $\hat a^\dagger$ as the creation operator for the instantaneous eigenstates of $\hat H_{LC}(t)$, whose energies are integer multiples of $\hbar (\partial_t \theta)(t)$.  

\section{Interaction Picture}
\label{interactionPicture}

Using the series of unitary transformations derived above, we now compute the operators $\hat \phi$ and $\hat q$ in the interaction picture with respect to Eq. \eqref{HLCtrans}. This serves two purposes. First, it will allow us calculate in the interaction picture the higher order terms $R(\hat \phi)$ of equation \eqref{R}, which we we dropped in order to the reach driven, quadratic Hamiltonian of equation \eqref{tHq2}. In the appendix we evaluate the size of these corrections in the rotated frame, and suggest a technique for minimizing their effect. Second, going to the interaction picture will allow us to write the original interaction $\hat H_I =  e\frac{ \xi}{d C_{\Sigma}} \hat q \hat z$ in terms of the effective annihilation operators of equation \eqref{a}, allowing for a straightforward calculation of the coherent coupling strength in the next section.

It is easier to first compute $\hat \phi$ in the interaction picture; the first two transformations are the displacements involved in linearization,
\beq
\hat \phi \rightarrow \hat \phi + \phi_c\,,
\eeq
while the third unitary $\hat U_3 = \exp(i \chi \hat \phi^2/\hbar)$ has no effect on $\hat \phi$. The final unitary $\hat U_4$ rescales $\hat \phi$ by the factor $r$, giving
\beq
\sqrt{\frac{\hbar}{2 C_\Sigma W}} r (\hat a + \hat a^\dagger) + \phi_c(t)\,,
\eeq
where we have used \eqref{a} to express $\hat \phi$ in terms of the annihilation operator $\hat a$. Finally, since in this rotated frame the circuit Hamiltonian is of the form $\hat H_{LC} = \hbar (\partial_t \theta) \l( \hat a^\dagger \hat a + 1/2\r)$, in the corresponding interaction picture $\hat a \rightarrow \hat a e^{-i \theta}$. The final form of $\hat \phi$ in the interaction picture is thus
\begin{align}
  \label{phiInt}
  \hat \phi_{int}(t) & =  \phi_c(t) + \sqrt{\frac{\hbar}{2 C_\Sigma W}} r (\hat a e^{-i \theta} + \hat a^\dagger e^{i \theta})\\
& = \phi_c(t) + \sqrt{\frac{\hbar}{2 C_\Sigma W}}(f^* \hat a + f \hat a^\dagger)\,, \label{phiint}
\end{align}
where we have used $f = r e^{i \theta}$. Using this expression for the interaction picture flux operator, in the appendix we bound the effect of the higher order corrections $R(\hat \phi)$ to the linearized Hamiltonian of equation \eqref{tHq}. A parameter of interest arising in this discussion is the relative size of the characteristic flux, which is set by
\beq
\label{smallflux}
\gamma \equiv \frac{1}{\tilde \phi_0}\sqrt{\frac{\hbar}{2 C_\Sigma W}}\,.
\eeq
Importantly, the final coupling strength is linearly proportional to this parameter, which we shall see limits the efficacy of our approach.

The derivation of $\hat q_{int}$ is analogous to that of $\hat \phi_{int}$. From the action of the four transformations $\hat U_1$ through $\hat U_4$,  $\hat q$ takes the form
\beq
q_c(t)  + \frac{1}{r} \hat q +  C_\Sigma (\partial_t r) \hat \phi\,.
\eeq
Expressing these operators in terms of $\hat a$ of equation \eqref{a} then going to the rotating frame $\hat a \rightarrow \hat a e^{-i \theta}$ gives
\beq
\hat q \rightarrow q_c(t) + \sqrt{\frac{\hbar C_\Sigma }{2 W }}\l( g(t) \hat a + g(t)^* \hat a^\dagger \r)\,,
\eeq
where $g(t) = (\partial_t r - i \frac{W}{r})e^{-i \theta} $. Using the Wronskian identity $W = r^2 \partial_t \theta$ of equation \eqref{Wronskian},  we see that $g(t) = \partial_t(r e^{-i \theta}) = \partial_t f^*$, so the final form of $\hat q$ in the interaction picture is
\beq
\label{qint}
\hat q_{int}(t) =  q_c(t) + \sqrt{\frac{\hbar C_\Sigma}{2 W}}\l(\partial_t f^* \hat a + \partial_t f\hat a^\dagger \r)\,.
\eeq
With this result, we may immediately compute the final ion-circuit interaction,
\begin{align}
\begin{split}
\hat H_{int}(t) &=  e\frac{ \xi}{d C_{\Sigma}} \hat q_{int}(t) \hat z_{int}(t)\\
&=    \xi \frac{e z_0}{C_\Sigma d} \sqrt{\frac{\hbar C_\Sigma }{2W }}\l( \partial_t f^* \hat a + \partial_t f \hat a^\dagger\r)\\
& \times ( \hat b e^{-i \omega_i t} + \hat b^\dagger e^{i \omega_i t}) \label{Hintfinal}\,,
\end{split}
\end{align}
where $\hat z_{int}(t) = z_0 ( \hat b e^{-i \omega_i t} + \hat b^\dagger e^{i \omega_i t}) $, with $z_0 = \sqrt{\frac{\hbar}{2 m \omega_i}}$ the characteristic displacement of the ion. Note that we have dropped the term $q_c(t) \hat z_{int}(t)$ by using the rotating wave approximation, since $q_c$ oscillates at characteristic frequency $\omega_d \gg \omega_i$. 

\section{Coupling strength}
In order to compute the effective coupling strength between the circuit and ion motion, we must first analyze the Fourier spectrum of the characteristic function $f$. We begin by expressing the interaction Hamiltonian of \eqref{Hintfinal} in terms of dimensionless factors,
\ba
\label{canonical}
 \nn\tilde H_{int}(u) &=& \frac{\hbar \omega_d}{4} \frac{z_0}{d} \xi \gamma \l( \partial_u f^* \hat a + \partial_u f \hat a^\dagger\r)\\
&& \times ( \hat b e^{-i 2 \omega_i u/\omega_d} + \hat b^\dagger e^{i2 \omega_i  u/\omega_d}) \label{Hintdimensionless}\,,
\ea
where we have used $\tilde \phi_0 = \hbar/(2 e)$ to get an expression in terms of the flux parameter $\gamma$ of equation~\eqref{smallflux}. Here we have rewritten $f$ in terms of the dimensionless parameter $u = \omega_d t/2$, to better match the canonical form of Mathieu's equation,
\beq
\l[\partial_u^2 + (A - 2 Q \cos 2u)\r] f = 0, \label{Mathieu}\,.
\eeq
Using the substitution $A = 4 (\omega_0/\omega_d)^2$, and $Q = -\eta a/2$, this equation is equivalent to \eqref{MathieuFirst}. By Floquet's Theorem, $f$ can be expressed as a quasi-periodic function
\beq
\label{fseries}
f(u) = e^{i \mu u} \sum_k c_k e^{2 i k u}\,,
\eeq
where the sum has period $u_0 = \pi$ corresponding to $t = 2 \pi/\omega_d$. Multiplying the cross terms of equation \eqref{Hintdimensionless}, we see that the coupling strength is proportional to the time-independent part of $ e^{i 2\omega_i u/ \omega_d } \partial_u f^* \hat a \hat b^\dagger$ (the only term to survive under the RWA). This constant is exactly proportional to the Fourier component of $f$ corresponding to the ion's motional frequency, which is specified by the resonance condition
\beq
\label{resonance}
 (\mu + 2 k) = 2\omega_i /\omega_d \,.
\eeq
Accounting for the derivative $\partial_u f^*$ in this expression, the coupling strength is
\beq
\label{coupling}
\hbar |\Omega| = \frac{\hbar \omega_d}{4} \frac{z_0}{d} \xi \gamma (\mu + 2 k)  |c_k| = \frac{\hbar \omega_i}{2} \frac{z_0}{d} \xi  \gamma  |c_k|\,,
\eeq
where in the second equality we used condition \eqref{resonance}. Note that although $\Omega$ is not explicitly dependent on the driving amplitude $\eta$ and frequency $\omega_d$, both $\gamma$ and $c_k$ are functions of these parameters since both are dependent on the characteristic function $f$.

Before we compute the effective coupling strength between the circuit and ion, we first rule out the presence of other resonances in equation~\eqref{canonical}. To do so we account for the ion's micromotion, which can be derived from its equation of motion in a time-dependent potential. Using calculations analogous to those in Sections~\ref{TDQHO} and~\ref{interactionPicture}, one may show that $\hat z_{int}(t) = z_0\l(\hat b e^{-i \omega_i t}h^*(t) + \mbox{h.c.} \r)$, where $h(t)$ is a periodic function whose fundamental frequency $\omega_{rf}$ matches the trapping potential's RF drive~\cite{Leibfried2003}. In the previous analysis we have approximated $h(t) = 1$, as this represents the largest Fourier coefficient of $h(t)$, while other coefficients correspond to frequencies displaced from $\omega_i$ by integer multiples of $\omega_{rf}$. Typically $\omega_{rf}$ ranges between $10-100$~MHz, in contrast to the $\omega_d \approx 1$~GHz frequency spacing of the charge oscillations (equations~\eqref{qint} and~\eqref{fseries}). Since the Fourier coefficients of both $h(t)$ and $f(t)$ are negligible at higher frequencies, the product $\hat z_{int}(t) \hat q_{int}(t)$ only has a resonance between $\hat a$ and $\hat b^\dagger$ at frequency $\omega_i$. By the same reasoning, the two-mode squeezing terms ($\hat a \hat b$ and $\hat a^\dagger \hat b^\dagger$) oscillate at frequency at least $2 \omega_i$ and may be dropped in the RWA. Likewise, no resonance exists between $\hat z_{int}(t)$ and the classical motion, since it $q_c(t)$ is a periodic function with fundamental frequency $\omega_d$. Thus, the only remaining term after making the rotating wave approximation on $\hat H_{int}(t)$ is the desired interaction Hamiltonian, $\hbar \Omega \,\hat a \hat b^\dagger + \mbox{h.c.}$. 

\begin{figure}
\includegraphics[width=.45\textwidth]{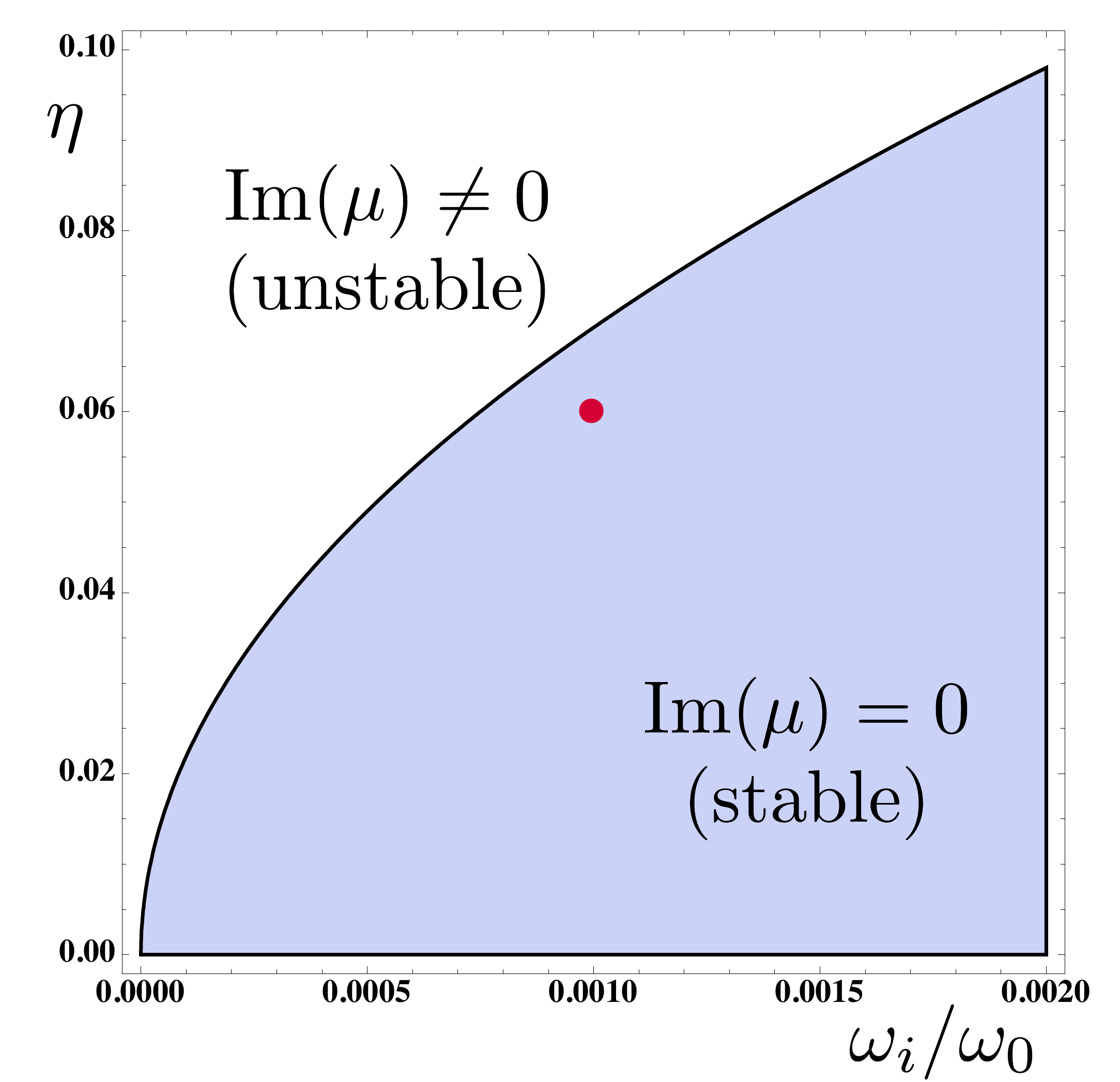} 
\caption{Stability diagram of Mathieu's equation in the resonant regime. The parameters $\omega_i/\omega_0$ and $\eta$ map to the canonical variables of equation \eqref{Mathieu} as $A = 4(1 - \omega_i/\omega_0)^2$ and $Q = -2\eta (1 - \omega_i/\omega_0)^2$.  The shaded region corresponds to stable quasi-periodic solutions, i.e. those with characteristic exponent $\mu$ having no imaginary part. The red dot at $\omega_i = 0.001\omega_0, \eta = 0.06$ corresponds to the driving parameters used in Fig.~\ref{fSmall}.}
\label{stability}
\end{figure}

With equation \eqref{coupling} we can now evaluate the strength of the coupling for a driving strategy similar to that of Ref.~\cite{Kielpinski2012}. Specifically, we set the drive frequency to be approximately the difference between the circuit's bare frequency and the ion's motional frequency: $\omega_d \approx \omega_0 - \omega_i$. Since the ion frequency $\omega_i \ll \omega_0$ is much smaller than the LC frequency, the drive frequency $\omega_d $ is nearly resonant with the LC circuit. This means that even a relatively small drive amplitude leads to a mathematical instability in the system. Indeed, for $\eta$ sufficiently large,  the characteristic exponent $\mu$ (equation~\eqref{fPeriod}) describing the quasi-periodic function $f$ gains an imaginary component, causing the interaction picture charge operator $\hat q_{int}(t)$ to diverge over time. This instability is alleviated in the presence environmental dissipation, as the system dynamics are then damped, though for simplicity we neglect these effects in our analysis.  As seen in Fig.~\ref{stability}, for $\omega_0 \approx \omega_d\gg \omega_i$ the boundary between mathematically stable and unstable regions is set by $\eta \lesssim 2 \sqrt{\omega_i/\omega_0}$. For near-resonant driving to be mathematically stable, the parameter $\eta$ must therefore be perturbatively small. In this $\eta\ll 1$ limit, the characteristic exponent is equal to $\mu = 2\omega_0/\omega_d + O(\eta^2)$ \cite{McLachlan1947}, so the solution of the frequency matching condition $\mu + 2 k = 2 \omega_i/\omega_d$ corresponds to $k = -1$. Using the relations $A = 4 (\omega_0/\omega_d)^2 \approx 4$ and $Q = - \eta A /2 \approx -2 \eta$, these parameters can be mapped to the canonical form of equation \eqref{Mathieu}. The coefficient $c_{-1} = Q/4 + O(Q^2) = - \eta/2 + O(\eta^2)$ is known from standard perturbative expansions \cite{McLachlan1947}. Substituting this value into equation \eqref{coupling}, we conclude that in the perturbative regime $\eta \lesssim 2 \sqrt{\omega_i/\omega_0}$, the coupling strength between LC and ion modes scales as
\ba
\hbar |\Omega| &= &\frac{\hbar \omega_i}{4} \l( \frac{z_0}{d} \xi \gamma \eta \r)\l(1 + O(\eta) \r)\nn \\
&\lesssim& \frac{\hbar \omega_i}{2} \l( \frac{z_0}{d} \xi \gamma \r) \sqrt{\frac{\omega_i}{\omega_0}}\l(1 + O(\sqrt{\omega_i/\omega_0})\r)\,. \label{coupling2}
\ea
We note that this driving scheme may not be optimal. When $\omega_d$ is set far from resonance and $\eta\sim 1$, the characteristic exponent $\mu$ can be stable and vary over a large range of values, allowing for the resonance condition \eqref{resonance} to be satisfied at stronger driving. Note that this also changes the Wronskian $W$ (which changes $\gamma \propto 1/\sqrt{W}$) in a non-linear way, so a general analysis for the best driving parameters is difficult. Strong driving may be infeasible for experimental reasons, since it may require too large a Josephson energy (which is proportional to $\beta>\eta$). 

Unfortunately, we find that the final effective coupling $\Omega$ is substantially weaker than in the proposal of Ref.~\cite{Kielpinski2012}. In that work, the ion's motional degree of freedom was also capacitively coupled to the circuit, but the characteristic equation of the circuit was instead modulated by a time varying capacitive element. The same driving scheme as above was used, leading to an effective coupling strength of the form,
\beq
\hbar \Omega_{cap} \sim \frac{e Q_0}{C_\Sigma} \l(\xi \frac{z_0}{d} \eta \r) \approx \hbar (2 \pi \times 60 \mbox{ kHz})\,.
\eeq
Here $Q_0 = \sqrt{\frac{\hbar C_\Sigma W}{2}}$ is the characteristic charge fluctuation in the driven circuit, and $\eta$ the relative amplitude of the time-varying capacitance, $C(t) = C_\Sigma ( 1 + \eta \sin(\omega_d t))$. For comparison, from equation \eqref{coupling2} we obtain a coupling rate $\Omega \simeq 1 \mbox{ Hz}$. Decoherence rates for these systems are expected to be on the order of $\rm{kHz}$, while leading order corrections from linearization scale as $\hbar \omega_i (\beta \gamma \omega_0/\omega_i)^2$, both of which make this approach infeasible as currently described\footnote{The presence of low frequency sidebands may also make the circuit somewhat more susceptible to $1/f$-like flux noise: Since the circuit's flux variable (Eq.~\eqref{phiInt}) gains a sideband at frequency $\omega_i$ scaling as $\sim \eta$ relative to its main Fourier component, we expect an additional low-frequency contribution to the decoherence rate. A simple Fermi's Golden Rule calculation suggests this should scale as $\eta^2 (\omega_i/\omega_d) \sim 1$, and is therefore comparable in magnitude to the undriven case.}. Note that we are making the same assumptions as Ref.~\cite{Kielpinski2012} about trap geometry ($\xi = 0.25, d = 25 \mu$m) and ion mass ($m \simeq 1.5 \times 10^{-26}$kg for $^9\mbox{Be}_+$), as well as ion motional frequency ($\omega_i \approx 2 \pi \times 1$MHz, so $z_0 = \sqrt{\frac{\hbar}{2 m \omega_i}} \approx 25 \mbox{ nm}$). Finally, the parameter $\gamma\sim 1.3$ is set by the circuit capacitance $C_\Sigma = 46 \mbox{ fF}$ and the Wronskian $W \approx \omega_0 = 2\pi \times 1 \mbox{ GHz}$.

As seen from the above comparison, when the ion-circuit interaction is capacitive (proportional to charge), modulating the circuit inductance at frequency $\omega_d \gg \omega_i$ is significantly less effective than modulating the capacitance. This results from the asymmetric dependence between charge and flux operators on the characteristic function $f$, as demonstrated by the interaction picture expressions for these operators (equations \eqref{phiint} and \eqref{qint}). While the flux scales with $f(t)$, charge is explicitly dependent on $\partial_t f(t)$. This means that the Fourier component of $\hat q_{int}(t)$ oscillating at frequency $\omega_i$ (and thus the time-independent component of $\hat H_{int}(t)$) picks up a factor of $\omega_i/\omega_d\ll 1$ compared to $\hat \phi_{int}(t)$. The opposite is true for capacitance modulation (for which the roles of $\hat q$ and $\hat \phi$ are reversed in transformations $\hat U_3$ and $\hat U_4$), which explains why it achieves a much larger effective coupling for a charge based interaction.


\section{Conclusions}

We have studied a technique to coherently couple a superconducting circuit to the motional mode of a trapped ion by careful variation of the circuit's inductance.  We describe a means of tuning the inductance (an external magnetic flux with a Josephson Junction) and describe an approximation mapping the circuit's non-linear Hamiltonian to that of a driven harmonic oscillator. Notwithstanding corrections to this approximation, the mismatch between the inductive driving and capacitive interaction is the major reason why the resulting coupling strength is impractically small. We confirm this in a direct comparison with a capacitive modulation scheme for a specific driving strategy, though the general form of the coupling suggests our conclusions hold for a broader class of strategies as well. Indeed, equation \eqref{coupling} holds for any choice of drive parameters $\omega_d, \eta$, and in fact can be applied more generally to any periodic modulation of inductance\footnote{Specifically, we may replace Mathieu's equation \eqref{MathieuFirst} with the more general Hill equation, $\partial_t^2 f + Q(t)f = 0$, where $Q(t)$ is any periodic function. The bare LC resonance frequency would then correspond to $\omega_0^2= \frac{1}{\tau}\int_0^\tau Q(t) \mbox{d}t$, where $Q(t + \tau) = Q(t)$.}. Conversely, our work suggests that for an inductive interaction (e.g. one based on mutual inductance between off-resonant circuits) an inductive modulation is the preferred approach.

\acknowledgements{The authors thank the following people for helpful discussions and comments concerning the work: Raymond Simmonds, Michael Foss-Feig, Hartmut H{\"a}ffner, David Kielpinski, Christopher Monroe, Dietrich Liebfried, and David Wineland. This research was supported by the U.S. Army Research Office Multidisciplinary University Research Initiative award W911NF0910406, and the NSF through the Physics Frontier Center at the Joint Quantum Institute.}


\bibliographystyle{apsrev}
\bibliography{bibli.bib}

\section{Appendix -- Linearization Procedure}
 The expression for $\hat \phi_{int}(t)$ allows us to evaluate the higher order corrections to the linearized Hamiltonian of equation \eqref{tHq2}. Since these corrections arise after the first two transformations ($(\hat \phi,\hat q)\rightarrow (\hat \phi + \phi_c, \hat q + q_c)$), they can be expressed as $\hat R_{int}(t) = R(\hat \phi_{int}(t) - \phi_c)$. Using relations \eqref{R} and $E_J = \beta \tilde \phi_0^2/L$ we obtain
\ba
\label{Rint}
\hat R_{int}(t) & =&\nonumber \sum_{k\geq3} \tilde \phi_0^k V^{(k)}_q|_{\phi_c(t)} \tilde \phi_0^{-k}\l(\hat \phi_{int}(t) - \phi_c(t) \r)^k/ k!\\
& =& \nonumber \frac{\tilde \phi_0^2}{L} \sum_{k\geq3} \beta c_k(t) \l(\sqrt{\frac{\hbar}{2 C_\Sigma W}}\frac{1}{\tilde \phi_0}\r)^k \l(\hat a f^* + \hat a^\dagger f \r)^k \frac{1}{k!}\\
& = & \nonumber\frac{\hbar}{ L C_\Sigma W} \sum_{k \geq 3} \beta c_k(t)  \gamma^{k-2} \l(\hat a f^* + \hat a^\dagger f \r)^k \frac{1}{k!}\,,
\ea 
where we have defined the characteristic flux parameter
\beq
\label{smallflux2}
\gamma \equiv \frac{1}{\tilde \phi_0}\sqrt{\frac{\hbar}{2 C_\Sigma W}}\,.
\eeq
Here $c_k(t) = E_J^{-1}\tilde \phi_0^k \frac{\partial^k V_q}{\partial \phi^k}|_{\phi = \phi_c}$ is defined according to equation \eqref{Vq}, giving
\ba
\nn \beta c_3(t) = &\beta \sin(\phi_c(t)/\tilde \phi_0)& = \sqrt{\beta^2 - \eta^2 \cos^2(\omega_d t)}\\
\nn \beta c_4(t) =& \beta \cos(\phi_c(t)/\tilde \phi_0)& = \eta \cos(\omega_d t)\\
\nn \beta c_k(t) = & -\beta \partial_x^k \cos(x) |_{x = \phi_c(t)/\tilde \phi_0} &,
\ea
where we have used $\beta \cos(\phi_c(t)/\tilde\phi_0) = \eta \cos(\omega_d t)$.

\begin{figure}

\includegraphics[width = 0.45\textwidth]{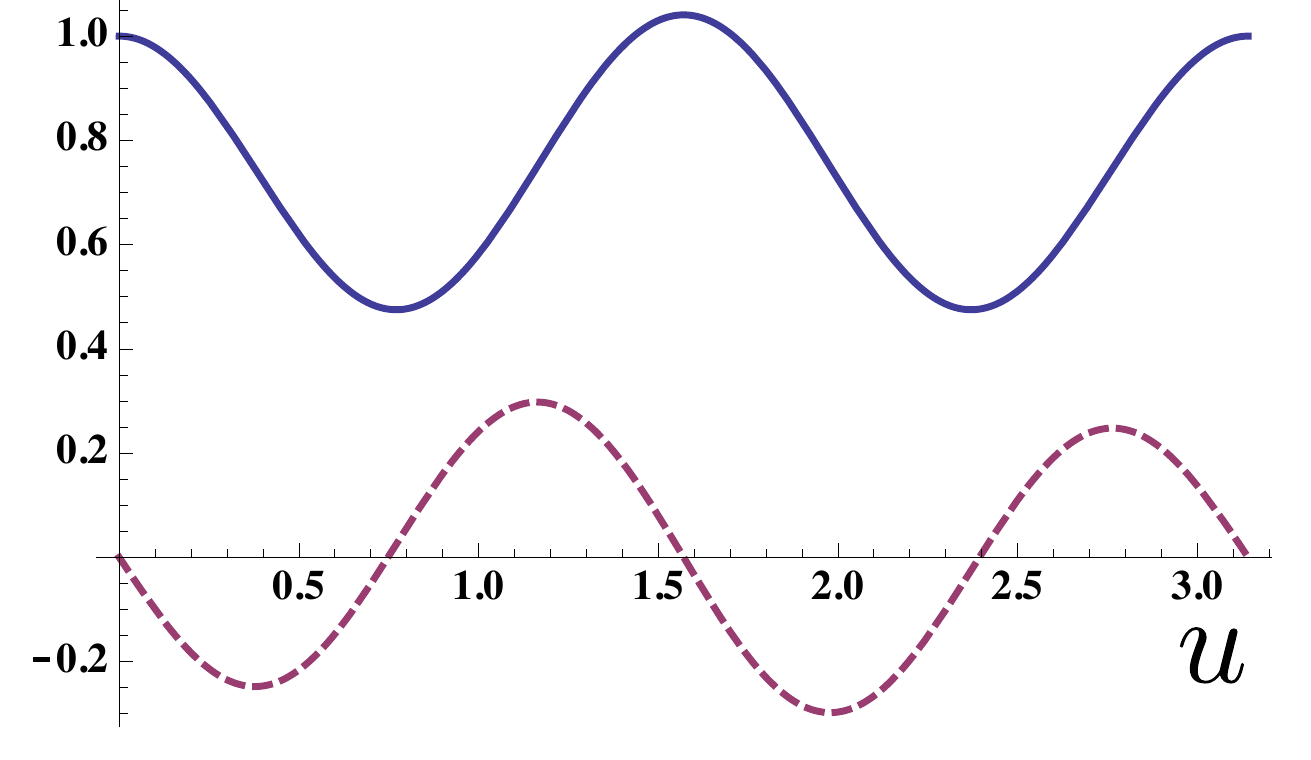} 
\caption{A plot of the periodic component of the characteristic function, $p(u) = e^{-i \mu u }f(u)$, in the weak driving regime. The blue, solid line and red, dashed lines represent the real and imaginary parts of $p(u)$, respectively. Because the driving frequency $\omega_d = 0.999\, \omega_0$ is nearly resonant with the bare LC frequency, even a small driving amplitude $\eta = 0.06$ causes significant deviations from the undriven case (for which $\mu u = 2 \pi \omega_0 t$ and $p(u) = 1$). }
\label{fSmall}
\end{figure}

To bound the error introduced by $\hat R_{int}(t)$ we begin by considering only the $k = 3$ contribution. Using $\omega_0^2 = 1/L C_\Sigma$ this term can be written as
\beq
\label{R3}
\hbar \omega_0 \frac{\omega_0}{W} \frac{\gamma \beta}{3!}  \sqrt{1 - (\eta/\beta)^2 \cos^2(\omega_d t)} \l(\hat a f^* + \hat a^\dagger f \r)^3\,.
\eeq
To analyze this term under the rotating wave approximation, we note that our coupling scheme is premised on giving $f$ a Fourier component matching the ion motional frequency, $\omega_i$. Specifically, in the driving scheme described in the text, the parameters $\omega_d\approx \omega_0-\omega_i$ and $\eta<\beta \ll 1$ are chosen such that $f$ has the form
\beq
f(t) = e^{i (\omega_d + \omega_i)t} \l( 1 + \frac{\eta}{2}\l( e^{i \omega_d t}/3 - e^{-i \omega_d t}\r) + ... \r)\,,
\eeq
where all other terms are of order $O(\eta^2)$ and correspond to frequencies $n \omega_d$ ($|n| \geq 2$). Thus the only slowly rotating term in equation \eqref{R3} is of order $\hbar \omega_0 \gamma \eta$  (since $W \approx \omega_0$ for these parameters) and rotates at frequency $\omega_i$. This slowly rotating part is the main contribution of \eqref{R3} for evolutions over a time scale $\sim 1/\omega_i$, and we may compute its effective size over this timescale by using a second-order Magnus expansion \cite{Magnus1954,Blanes2009},
\beq
\label{size}
\hat R_{int}(t) \sim \hbar \omega_i \l(\gamma \beta \frac{\omega_0}{\omega_i}\r)^2\,.
\eeq
Using a similar procedure, we can derive analogous estimates for the higher ($k >3$) order terms as well. But because these terms pick up extra factors of $\gamma$, equation \eqref{size} represents the overall scaling of all higher order terms. This is true when the scale of $|f|$ is on the order of $\sim 1$ (as in Fig.~\ref{fSmall}), the characteristic flux is small $(\gamma \lesssim 1)$, and there are only small fluctuations above the classical solution, $\Mean {\hat a^\dagger \hat a} \sim 1$.

One technique to reduce the effective size of corrections $\hat R_{int}(t)$ is to replace the single Josephson junction with a linear array of these elements. In the simplest approximation\cite{Kaplunenko2004}, using a stack of $N$ junctions we get that the Josephson Hamiltonian contribution is transformed as
\beq
-E_J \cos\l(\frac{\hat \phi}{\tilde \phi_0}\r) \rightarrow -N E_J  \cos\l(\frac{\hat \phi}{N\tilde \phi_0}\r)
\eeq
In terms of the parameters in the definition of $\hat R_{int}(t)$, this corresponds to the map $\beta\rightarrow \beta/N$, $\tilde \phi_0 \rightarrow N \tilde \phi_0 $ (or equivalently $\gamma\rightarrow \gamma/N$).  From equation \eqref{size}, this rescales the leading order contribution of $\hat R_{int}(t)$ by a factor of $1/N^4$. For the resonance ratio $\omega_0/\omega_i = 1000$, an array of $N \sim 100$ junctions should suffice to limit the effect of all higher order corrections.

\end{document}